\documentclass[superscriptaddress,pre,twocolumn,amsmath,amssymb]{revtex4} 
\usepackage{graphicx,natbib}
\usepackage[caption=false]{subfig}

\newcommand{\vect}[1]{\mathbf{#1}} 
\newcommand{\RM }[1]{\mathrm{#1}}
\newcommand{\ave}[1]{ {\langle {#1} \rangle} }

\def\br{\vect{r}}

\def\kB{ k_{\RM{B}} }

\def\s2{{s_2}}
\def\Is2{{I_{\s2}}}

\def\Visc0{{\eta_{0}}}

\def\sx{{ s^{\RM{ex}} }}
\def\exp{{ \RM{exp} }}

\begin{document}

\title{Anomalous structure and dynamics of the Gaussian-core fluid}

\author{William P. Krekelberg}
\email{wpkrekelberg@gmail.com} 
\affiliation{Department of Chemical
  Engineering, University of
  Texas at Austin, Austin, TX 78712.}  

\author{Tanuj Kumar}
\affiliation{Department of Chemical
  Engineering, University of
  Texas at Austin, Austin, TX 78712.}  

\author{Jeetain Mittal}
\email{jeetain@helix.nih.gov} 
\affiliation{Laboratory of Chemical
  Physics, NIDDK, National Institutes of Health, Bethesda, MD
  20892-0520. }  

\author{Jeffrey R. Errington}
\email{jerring@buffalo.edu}
\affiliation{Department of Chemical and Biological Engineering, 
  University at Buffalo, The State University of New York, 
  Buffalo, New York 14260-4200, USA}

\author{Thomas M. Truskett}
\email{truskett@che.utexas.edu}
\thanks{Corresponding Author} 
\affiliation{Department of Chemical
  Engineering, University of
  Texas at Austin, Austin, TX 78712.}  
 \affiliation{Institute for Theoretical Chemistry, 
   University of Texas at Austin, Austin, TX 78712.}

\begin{abstract}

It is known that there are thermodynamic states for which the 
Gaussian-core fluid displays anomalous properties such as expansion upon 
isobaric cooling (density anomaly) and increased single-particle
mobility upon isothermal compression (self-diffusivity anomaly).
Here, we investigate how
temperature and density affect its short-range translational structural
order, as characterized by the two-body excess entropy.  
We find that there is a wide range of conditions for which the 
short-range translational order of the Gaussian-core fluid 
decreases upon isothermal compression 
(structural order anomaly).  As we show, the origin of the
structural anomaly is qualitatively similar to that of 
other anomalous fluids (e.g., water or
colloids with short-range attractions) and is connected to how
compression affects static correlations at different length scales.  
Interestingly, we find that the 
self-diffusivity of the Gaussian-core fluid 
obeys a scaling relationship with 
the two-body excess entropy that is very similar 
to the one observed for a variety of simple liquids.  One consequence of this
relationship is that the
state points for which structural, self-diffusivity, and 
density anomalies of the Gaussian-core fluid 
occur appear as cascading regions on the
temperature-density plane, a phenomenon observed earlier for models
of waterlike fluids.  There are, however, key differences between the
anomalies of Gaussian-core and waterlike fluids, and we discuss how
those can be qualitatively understood
by considering the respective interparticle potentials of these models.   
Finally, we note that the self-diffusivity of the Gaussian-core 
fluid obeys different scaling laws depending on whether the two-body or total
excess entropy is considered.  This finding, which deserves more
comprehensive future study, appears to underscore the
significance of higher-body correlations for the behavior of fluids
with bounded interactions.   
\end{abstract}
\maketitle

Model pair potentials that describe the interactions between particles
of simple atomic and molecular liquids generally have steeply
repulsive, short-range components that diverge as two particles
approach one another, qualitatively capturing the increase in energy
that accompanies overlap of their electronic clouds.  However, the
effective pair interactions between the centers of mass of two
polymer chains in solution (or molecular aggregates such as micelles)
are often more accurately represented by softer, bounded
potentials~\cite{Flory1950Statistical-Mec}, since they describe
``particles'' that are inherently interpenetrable.  A simple example
of a bounded interaction is given by the Gaussian-core (GC)
model \cite{Stillinger1976Phase-transitio}.  It is defined by a pair
potential of the following form, $\phi(r)=\epsilon
\,\RM{exp}[-(r/\sigma)^2]$, where $r$ is the distance between particle
centers, and $\epsilon$ and $\sigma$ characterize the amplitude and
width of the interaction profile, respectively.  Understanding the
differences between the collective properties of penetrable particles
with bounded effective potentials and those with steeply repulsive
interactions is a basic problem in the study of soft condensed matter
(see, e.g., \cite{Lang2000Fluid-and-solid,Likos2008Cluster-forming,Zachary2008Gaussian-core-m}). 


The behavior of the GC fluid is indeed unusual when compared to most
atomic or molecular fluids.  For example, there are state points for
which its thermal expansion coefficient is negative (density
anomaly)~\cite{Stillinger1997Negative-therma} and for which its
single-particle dynamics increase upon isothermal compression
(diffusivity anomaly)~\cite{Mausbach2006Static-and-dyna}.  At
sufficiently low temperature, the fluid also exhibits re-entrant
melting behavior
\cite{Stillinger1976Phase-transitio,Lang2000Fluid-and-solid,Prestipino2005Phase-diagram-o}.
Anomalous changes of self-diffusivity with density or temperature are
known to occur in other model systems that qualitatively mimic either
the behaviors of (i) liquid water or (ii) suspensions of colloids with
short-range interparticle attractions, respectively.  A series of
theoretical studies have shown that those changes for both systems of
type
(i)~\cite{Errington2001Relationship-be,Shell2002Molecular-struc,Truskett2002A-Simple-Statis,Kumar2005Static-and-dyna,Esposito2006Entropy-based-m,Netz2006Thermodynamic-a,Xu2006Thermodynamics-,Mittal2006Quantitative-Li,Errington2006Excess-entropy-,Sharma2006Entropy-diffusi,Oliveira2007Interplay-betwe,Yan2007Structure-of-th,Szortyka2007Diffusion-anoma,Oliveira2008Waterlike-hiera,Krekelberg2008Structural-anom,Yan2008Relation-of-wat}
and type
(ii)~\cite{Mittal2006Quantitative-Li,Krekelberg2007How-short-range,Krekelberg2008Structural-anom}
are strongly correlated to anomalous trends in the short-range
translational structural order of the respective fluids.  Moreover, an
analysis of recent Brownian dynamics simulations of the GC fluid \cite{Wensink2008Long-time-self-}
 suggested a
qualitative (but not quantitative) connection between the anomalous
density dependencies of the self-diffusivity and the pair correlation
function.  Here, we carry out an extensive molecular simulation study
of the GC fluid to investigate whether a quantitative link between its
dynamics and its short-range translational order can be established.

To compute the properties of the GC model, we perform molecular
dynamics simulations in the microconanical ensemble using $N=1000$
particles in a cubic cell with periodic boundary conditions.  
For
simplicity of notation, we implicitly non-dimensionalize all
quantities by appropriate combinations of the characteristic length
scale $l_\RM{c}=\sigma$ and time scale $\tau_{c}=\sqrt{m
  \sigma^2/\epsilon}$, where $m$ is the mass of a particle.  We
integrate the equations of motion using the velocity-Verlet method
\cite{Allen1987Computer-Simula} with time step $\delta t=0.05$, and we
cut the GC potential at $r_{\RM{cut}}=3.2$.  We calculate
self-diffusivity by fitting the long-time ($t\gg 1$) behavior of the
mean-squared displacement $\ave{\delta \br^2(t)}$ to the Einstein
formula $6Dt=\ave{\delta \br^2(t)}$. We investigate temperatures
between $T=0.01$ and $1.5$ and number densities between $\rho=0.01$
and $2.0$ \cite{note_temp_dens}, parameters that span the part of
phase space where the aforementioned anomalous behavior is known to
occur for the GC fluid~\cite{Mausbach2006Static-and-dyna}.

\begin{figure*}[t]
  \centering
  \includegraphics[keepaspectratio,clip]{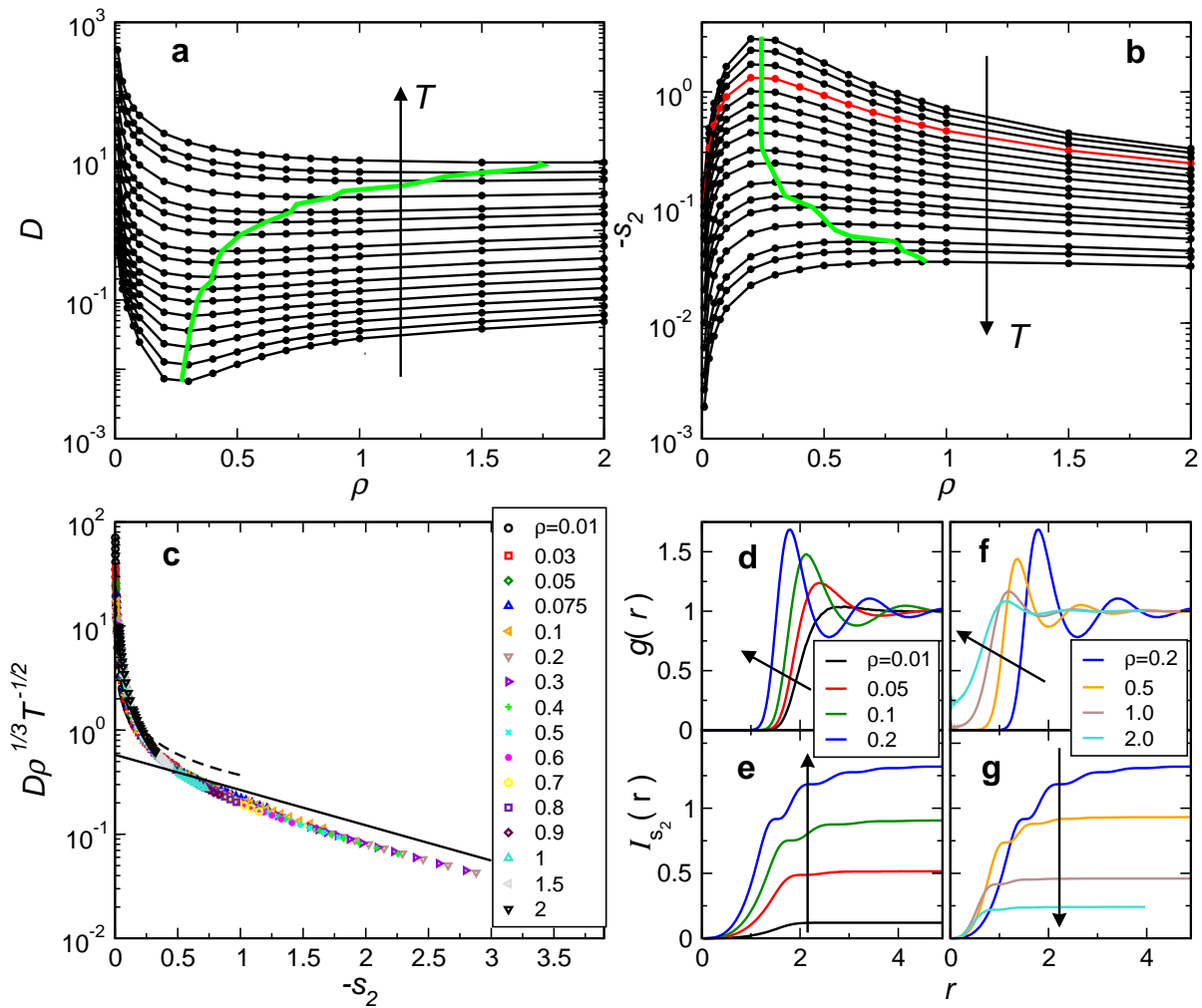}
  \subfloat{\label{fig:diff_vs_rho}}    
  \subfloat{\label{fig:s2_vs_rho}}
  \subfloat{\label{fig:drosen_s2}}
  \subfloat{\label{fig:s2fig_gofr_up}}
  \subfloat{\label{fig:s2fig_is2_up}}
  \subfloat{\label{fig:s2fig_gofr_down}}
  \subfloat{\label{fig:s2fig_is2_down}}
  
  \caption{ (Color online)  \textbf{(a)} Self-diffusivity $D$ and \textbf{(b)}
    structural order metric $-s_2$ versus density $\rho$ at temperatures
    $T=0.01,0.013,0.018,0.025,0.035,0.05,0.07,0.1,0.15,0.2,0.3,0.4,0.5,0.7,1.0,1.2$, and $1.5$ for the GC fluid.
    Arrows indicate increasing temperature. Solid green lines
    in \textbf{(a)} and \textbf{(b)} correspond to minima in $D$ and
    maximum in $-s_2$ , respectively. \textbf{(c)} Rosenfeld scaled diffusivity
    versus $-s_2$ at all densities and temperatures studied.    
The solid line
    represents the empirical Rosenfeld scaling law $0.58\exp(0.78 s_2)$ and the
    dashed curve represents the low density behavior expected for soft spheres
    $0.37(-s_2)^{-2/3}$.  (Lower right panel) Pair correlation
    function $g(r)$ and cumulative order integral
    $\Is2(r)$ along the isotherm $T=0.025$ [red curve in \textbf{(b)}]
    for two density ranges [\textbf{(d)} and \textbf{(e)}]
    $\rho\leq0.2$ [below the maximum in $-s_2(\rho)$] and
    [\textbf{(f)} and \textbf{(g)}] $\rho\geq 0.2$ [above the maximum
    in $-s_2(\rho)$]. In \textbf{(d-g)}, arrows indicate increasing $\rho$.  }
  \label{fig:diffusion_structure}

\end{figure*}

Figure~\ref{fig:diff_vs_rho} displays the self-diffusivity $D$ of the
GC fluid versus density $\rho$ for several isotherms.  Like in simpler
fluids, compressing the low density GC fluid initially decreases $D$.
However, as noted previously \cite{Mausbach2006Static-and-dyna},
compression can eventually lead to an anomalous increase in $D$ at
sufficiently high particle density.  In fact, for a given $T$, one can
view the density at which the minimum in $D$ occurs as a boundary
between regions of ``normal'' and anomalous dynamic behavior.  This
boundary is approximately indicated by the green line in
Fig.~\ref{fig:diff_vs_rho}~\cite{note_min_max}.

To address the question of whether this anomalous dynamic behavior is
related to structural order, we first investigate the behavior of the
two-body contribution to the excess entropy (relative to an ideal gas
at the same $T$ and $\rho$), defined as
\cite{Nettleton1958Expression-in-T,Baranyai1989Direct-entropy-}
\begin{equation}
  \label{eq:s2}
  s_2\equiv -2\pi\rho \int_0^\infty r^2 \{ g(r) \RM{ln}\, g(r)
  -[g(r)-1]\} dr,
\end{equation}
where $g(r)$ is the pair correlation function (PCF).  The quantity
$-s_2$ effectively characterizes the degree of pair translational order
present in the fluid \cite{Truskett2000Towards-a-quant}.
Figure~\ref{fig:s2_vs_rho} displays $-s_2$ versus $\rho$ for several
isotherms.  Like in simple fluids, compressing the low density GC
fluid initially increases structural order.  However, further
increasing the density eventually leads to a qualitative change in
behavior.  Specifically, the GC fluid begins to lose structural order
upon compression at sufficiently high density, a trend that does not
generally occur in atomic
fluids~\cite{Baranyai1989Direct-entropy-,Truskett2000Towards-a-quant}.
Thus, for a particular $T$, the density corresponding to the maximum
in $-s_2$ can similarly be viewed as a boundary between state points
of ``normal'' and anomalous structural order.  This boundary is approximately
indicated by the green line in Fig.~\ref{fig:s2_vs_rho} \cite{note_min_max}.

Together, Figures~\ref{fig:diff_vs_rho} and \subref*{fig:s2_vs_rho}
suggest a strong correlation between dynamics and structural order for
the GC fluid.  Given the recent results from computer simulations
\cite{Rosenfeld1977Relation-betwee,Dzugutov1996A-univeral-scal,Rosenfeld1999A-quasi-univers,Mittal2006Quantitative-Li,Mittal2006Thermodynamics-,Mittal2006Relationship-be,Sharma2006Entropy-diffusi,Mittal2007Relationships-b,Krekelberg2007How-short-range,Mittal2007Does-confining-,Mittal2007Confinement-ent,Goel2008Tuning-Density-,Agarwal2007Ionic-melts-wit,Yan2008Relation-of-wat}
and experiments
\cite{Abramson2008Viscosity-of-ni,Abramson2007Viscosity-of-wa} which
show a clear connection between dynamics and excess entropy for a wide
variety of systems, this correlation may not be particularly
surprising.  In fact, an approximate scaling law relating transport
coefficients and excess entropy per particle, $\sx$, was first noted
by \citet{Rosenfeld1999A-quasi-univers,Rosenfeld1977Relation-betwee}.
Specifically, Rosenfeld observed that the following scaled form of
self-diffusivity, $D_R=D\rho^{1/3}T^{-1/2}$, is a quasi-universal
function of $\sx$ for a variety of models of simple atomic fluids.  In
other words, the dependencies of $D$ on $T$ and $\rho$ can essentially
be reduced to how these variables affect $\sx$.  For simple liquid
systems at low to intermediate densities, most of the excess entropy
($\sx$) comes from the two-body contribution
($s_2$)~\cite{Baranyai1989Direct-entropy-}, and thus the
quasi-universal excess entropy scaling indicates that their transport
coefficients are closely linked to $g(r)$ (see also, e.g., \cite{Samanta2001Universal-Scali}).

For atomic fluids, the scaling law relating $D$ and $\sx$ closely
tracks different functional forms as $-\sx$ goes from lower to higher values.  
For very low $-\sx$ (i.e., a dilute gas), the following power-law form
follows from Enskog theory for $D$ 
and a second-virial approximation for $\sx$ (see~\cite{Rosenfeld1999A-quasi-univers}),
\begin{subequations}
  \label{eq:diff_rosen}
  \begin{equation}
    \label{eq:diff_rosen_low}
    D\rho^{1/3}T^{-1/2}\approx \alpha(-\sx)^{\beta},
  \end{equation}
  where $\alpha$, $\beta$ are constants.  At higher $-\sx$ (e.g., a
  fluid near the freezing transition), the following exponential
  relation has been empirically
  observed~\cite{Rosenfeld1999A-quasi-univers}
  \begin{equation}
    \label{eq:diff_rosen_high}
    D\rho^{1/3} T^{-1/2} \approx A\, \exp[B\, \sx],
  \end{equation}
  where $A$ and $B$ are constants.    
\end{subequations}

Figure~\ref{fig:drosen_s2} displays the scaled self-diffusion
coefficient $D_R$ versus the two-body estimate of the excess entropy
$-s_2$.  Note that the dependencies of the Rosenfeld self-diffusivity
$D_R$ on $T$ and $\rho$ for the GC fluid also collapse onto a single
master curve when plotted versus $-s_2$.  For comparison,
Eq.~\eqref{eq:diff_rosen} with parameters that fit the simulation data
for a variety of soft-sphere model systems ($\alpha=0.37$,
$\beta=-2/3$ \cite{Rosenfeld1999A-quasi-univers} and $A=0.58$,
$B=0.78$ \cite{Rosenfeld1977Relation-betwee}) is also shown in
Fig.~\ref{fig:drosen_s2}.  The main point to note here is that,
despite small quantitative differences, the form of the relationship
between the diffusivity and excess entropy is virtually the same,
whether one considers the anomalous GC fluid or a model for a simple
atomic fluid.  In other words, despite the anomalous dependency of $D$
upon $\rho$ for the GC fluid, the relationship between $D$ and $s_2$
is perfectly normal. 
\begin{figure*}[t]
  \centering
  \mbox{
    \includegraphics[keepaspectratio,clip]{anomalies.eps}
    \subfloat{\label{fig:anomalies-data}}
    \includegraphics[keepaspectratio,clip]{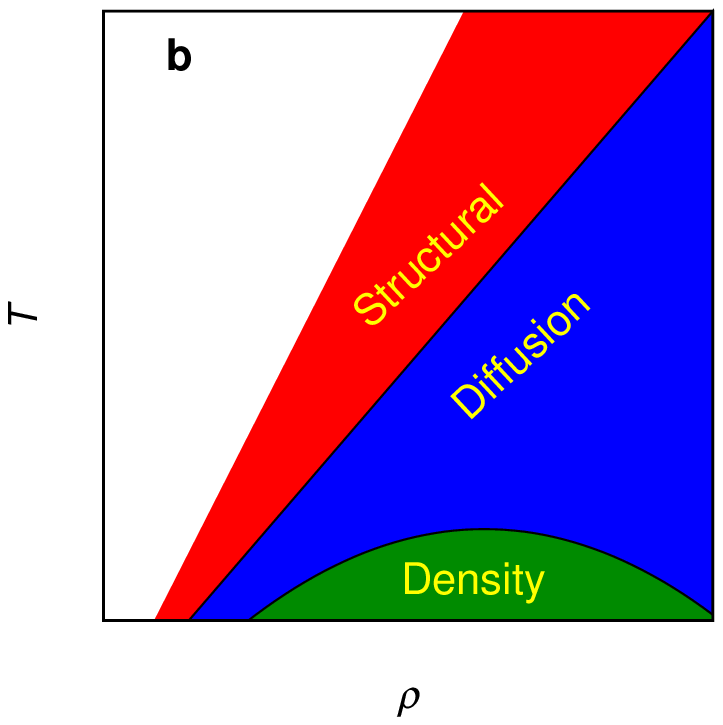}
    \subfloat{\label{fig:anomalies-schematic}}
  }
  \caption{(Color online)  \textbf{(a)} Boundaries of the structural, diffusivity,
    and density anomalies in the temperature-density plane, as
    calculated from simulation data. The inset shows the boundary of
    the region of density
    anomalies on an expanded scale.  The symbols are the locations of
    the extrema, and arrows are in the direction of the anomalous
    regions. \textbf{(b)} Schematic
    representation of anomalous regions.}
  \label{fig:anomalies_all}
\end{figure*}

Given this basic link between anomalous dynamics and structure, we
now investigate the molecular origins of the trends in
$-s_2$.  To this end, we study the PCF and the related cumulative order
integral $\Is2(r)$ defined as~\cite{Krekelberg2007How-short-range}
\begin{equation}
  \label{eq:Is2}
  I_{s_2}(r)\equiv 2\pi \rho\int_0^r r'^{2} \{g(r') \RM{ln}g(r') - [g(r')-1]\} dr'. 
\end{equation}
Note that one recovers $-s_2$ from this integral in the large $r$
limit.  The quantity $I_{s_2}(r)$ characterizes the average amount of
translational ordering on length scales smaller than $r$ surrounding a
particle. Figs.~\ref{fig:s2fig_gofr_up} - \subref*{fig:s2fig_is2_down}
show the behavior of $g(r)$ and $I_{s_2}(r)$ for selected state points
along the low-temperature $T=0.025$ isotherm (red curve in
Fig.~\ref{fig:s2_vs_rho}).  They divide the behavior into two
qualitatively different regions: (1) the normal increase of $-s_2$
with $\rho$ at low densities [$\rho\leq 0.2$,
Figs.~\ref{fig:s2fig_gofr_up} and \subref*{fig:s2fig_is2_up}] and (2)
the anomalous decrease of $-s_2$ with $\rho$ at high densities
[$\rho\geq 0.2$, Figs.~\ref{fig:s2fig_gofr_down} and
\subref*{fig:s2fig_is2_down}].

Inspection of Fig.~\ref{fig:s2fig_gofr_up} shows that
compression of the low-density GC fluid leads to an increased degree of
translational order, reflected by the formation of distinct first, second, and
more distant coordination shells in $g(r)$.  
This increase in ``packing'' order around the particles (see Fig.~\ref{fig:s2fig_is2_up}) is
also observed in hard-sphere fluids and simple atomic systems, and it
is a reflection of the fact that local 
interparticle correlations necessarily build up as density increases 
in order for particles to avoid overlap with one another.  At
sufficiently low pressures (low densities) and low temperature, the 
effective repulsive 
``core'' of the GC potential acts like a hard-sphere interaction, and
that is the primary reason that 
the structural trends of the GC fluid under these conditions 
mimic those of simple atomic systems.     

Further compression of the GC fluid [region (2), $\rho\geq 0.2$] leads
to an anomalous decrease in structural order
[Fig.~\ref{fig:s2_vs_rho}].  In this higher pressure, higher density 
region of phase space, the 
repulsion of the GC interparticle potential is too soft to 
effectively prevent interparticle
overlap.  In fact, because the system only pays a finite potential 
energy penalty for each overlap, and since avoiding overlaps carries a
substantial entropic penalty, the system evolves toward 
a more uniform average structure at sufficiently high density.  
In the limit of infinitely high 
density, the particle centers of the GC fluid will adopt a Poisson 
distribution (i.e., a high-density ideal gas)
\cite{Lang2000Fluid-and-solid}. 
As can be seen in Figure~\ref{fig:s2fig_gofr_down}, these compression
induced changes manifest as a shifting in, 
flattening, and broadening
of the coordination shells in $g(r)$.
Figure~\ref{fig:s2fig_is2_down} shows that while finite penetrability gives rise to a small increase in
order at small $r$ (inside the core), the overall behavior is
dominated by the corresponding disordering that occurs at larger $r$. 

The microscopic origins of the structural behavior of the GC 
fluid share a qualitative similiarity with those of the other anomalous fluids discussed 
earlier (liquid water and suspensions of colloids with short-range
attractions)~\cite{Krekelberg2008Structural-anom}. In the case
of models that capture waterlike behavior, compression of the low
temperature fluid can result in an anomalous translational 
disordering~\cite{Errington2001Relationship-be,Shell2002Molecular-struc,Truskett2002A-Simple-Statis,Yan2005Structural-Orde,Yan2006Family-of-tunab,Oliveira2006Structural-anom,Mittal2006Quantitative-Li,Mittal2006Relationship-be,Sharma2006Entropy-diffusi,Oliveira2007Interplay-betwe,Agarwal2007Ionic-melts-wit,Agarwal2007Waterlike-Struc,Yan2007Structure-of-th,Oliveira2008Waterlike-hiera}.
This decrease in structural order is primarily due to a shifting in,
flattening, and broadening of 
the second coordination shell \cite{Yan2007Structure-of-th,Krekelberg2008Structural-anom}), which is accompanied by the penetration
of a fifth water molecule into
the periphery of the first shell of otherwise tetrahedrally-coordinated
molecules (see, e.g., \cite{Yan2007Structure-of-th}).  In the
case of colloids with short-range attractions, increasing the
strength of the attractive interactions (relative to $k_{\mathrm B} T$) of the 
concentrated fluid can also give rise to an anomalous
disordering~\cite{Mittal2006Quantitative-Li,Krekelberg2007How-short-range}.
This disordering is due to the fact that the short-range attractions
favor near contact configurations of the particles.  This disrupts the
structure of the second and more distant coordination shells 
that otherwise naturally form when structure
is determined primarily 
by the repulsive interactions (i.e., as in hard-sphere or
simple atomic fluids) \cite{Krekelberg2008Structural-anom}.  Thus, in
each of the above systems, there are two effects that are
linked: (i) an increase
in correlations at short length scales and (ii) an associated 
decrease in structural order at larger length scales.  The latter
effect dominates in these fluids, thus giving rise to their 
structurally anomalous behavior.

It is also interesting to ask about the
relative placement of the boundaries for structural, diffusivity, and
density anomalies on the temperature-density plane.  
Fig.~\ref{fig:anomalies-data} shows those boundaries for the GC
fluid as determined from our simulation data, and 
Fig.~\ref{fig:anomalies-schematic} provides a idealized schematic
based on that data.  
The most striking feature is that the state points which exhibit the
density anomaly (negative thermal expansion coefficient) 
are a subset of those associated with the diffusivity
anomaly, which, in turn, are a subset of those associated with the
structural anomaly.  This ``cascade of anomalies'' is qualitatively
similar to that originally noticed for a molecular model of water by
Errington and Debenedetti~\cite{Errington2001Relationship-be}, and
more recently observed in a wide class of model systems that exhibit
waterlike
behavior~\cite{Yan2006Family-of-tunab,Oliveira2006Structural-anom,Oliveira2006Thermodynamic-a,Errington2006Excess-entropy-,Shell2002Molecular-struc,Oliveira2006Structural-anom,Yan2007Structure-of-th,Truskett2002A-Simple-Statis,Oliveira2007Interplay-betwe,Sharma2006Entropy-diffusi,Oliveira2008Waterlike-hiera,Kumar2005Static-and-dyna,Esposito2006Entropy-based-m,Netz2006Thermodynamic-a,Xu2006Thermodynamics-,Szortyka2007Diffusion-anoma}.  
One qualitative difference between the GC fluid
and water is that, while water returns to normal structural and
diffusivity trends at sufficiently high density, the GC fluid does
not.  This difference follows from the fact that water has steeply
repulsive short-range interactions, which dominate its physics at
sufficiently high densities (pressures).  The repulsive core is, of
course, absent in bounded potentials like the GC model, and thus those
systems never ``return'' to hard-sphere-like structural and dynamic 
behavior at high density.  However, like water, the GC fluid does return to
normal (i.e., positive) thermal expansion behavior at sufficiently
high density.  This is consistent with the fact that the GC fluid's 
equation of state approaches a mean-field description in that 
limit~\cite{Lang2000Fluid-and-solid}.  
Finally, we note that \citet{Oliveira2006Thermodynamic-a} have created a
simple model for water with a spherically-symmetric pair
potential represented by a sum of a Lennard-Jones contribution (which
contains the necessary repulsive core) and 
a Gaussian-core interaction.  Given the discussion above, it should
not be surprising that this generic 
type of hybrid model can qualitatively reproduce 
water's density, diffusivity, and structural
anomalies~\cite{Yan2005Structural-Orde,Yan2006Family-of-tunab,Oliveira2006Structural-anom,Oliveira2008Waterlike-hiera}.

It is also interesting to note that the state points where models of
colloidal fluids with short-range attractions show diffusivity
anomalies are similarly a subset of the state points where they
exhibit structural anomalies~\cite{Krekelberg2007How-short-range}.  
As has now been discussed extensively for models of
water~\cite{Errington2006Excess-entropy-} and for models of colloids with
short-range attractions~\cite{Krekelberg2007How-short-range}, the
``cascade of anomalies'' has been shown to be qualitatively consistent
with the behavior
of the excess entropy,
the empirical relationship between 
diffusivity and excess entropy, and the rigorous thermodynamic link
between the excess entropy and the thermal expansion coefficient.

\begin{figure}[t]
  \centering
  \includegraphics[keepaspectratio,clip]{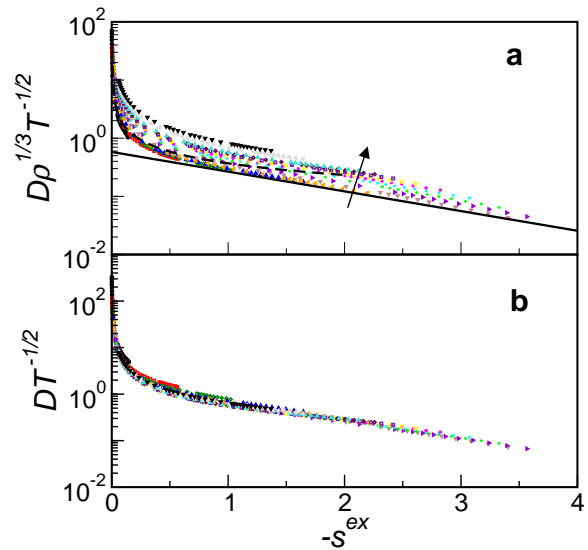}
  \subfloat{\label{fig:diff_sx_rosen}}
  \subfloat{\label{fig:diff_sx_thalf}}
  \caption{ \textbf{(a)} Rosenfeld scaled diffusivity versus excess
  entropy $-\sx$.  The solid and dashed curves, as well as the
  symbols, are the same as those
  in Fig.~\ref{fig:drosen_s2}.  The arrow is in the direction of
  increasing density. \textbf{(b)} Temperature scaled diffusivity
  $DT^{-1/2}$ verses $-\sx$.  The data represent all densities and
  temperatures considered in this study.}
  \label{fig:sxfig}
\end{figure}

Thus far, we have focused exclusively on the two-body contribution to
the excess entropy ($s_2$) of the GC fluid rather than the excess
entropy itself ($\sx$).  The rationale for doing so is simple.  The
two-body quantity is a particularly convenient structural measure to
compute and analyze, given its straightforward connection to $g(r)$.
Furthermore, as discussed above, $s_2$ is known to closely approximate
$\sx$ for simple fluids at low to intermediate
density~\cite{Baranyai1989Direct-entropy-}.  However, the GC fluid has
structural behavior that is very different from that of simple atomic
fluids.  In fact, it is known that 
there are significant differences between $s_2$ and $\sx$
for this system, due to non-canceling effects from higher-order
static correlations (3-body, 4-body, etc.)
\cite{Giaquinta2005Re-entrant-Melt}.  As an example, 
along the $T=0.025$ isotherm highlighted above, the ratio $\sx / s_2$
evaluates to approximately $1.7$ at $\rho=0.3$ (where $-\sx$ reaches a
maximum) and $3.0$ at $\rho = 1.0$.  
Thus, as a final point, we
present in Fig.~\ref{fig:diff_sx_rosen} the relationship between the
Rosenfeld diffusivity, $D_R=D\rho^{1/3}T^{-1/2}$, and $-\sx$ for all
state points considered in this study of the GC fluid.  We calculate
$\sx$ in this study using grand-canonical transition-matrix Monte
Carlo (GC-TMMC)
simulations~\cite{Errington2003Direct-calculat,Errington2003Evaluating-surf,Errington2006Excess-entropy-,note_tmmc}.

Note that, for a given density, the relationship
between $D_R$ and $-\sx$ is qualitatively similar to that of atomic
fluids and hence also to the relationship between $D_R$ and $-s_2$ for the
GC fluid shown in Fig.~\ref{fig:drosen_s2}.  However, 
the $D_R$ data at different densities do
not collapse onto a single curve when plotted versus $-\sx$ (contrast
with Fig.~\ref{fig:drosen_s2}).  As expected, the
low-density data are indeed accurately represented by
Eq.~\eqref{eq:diff_rosen} with parameters that describe soft-sphere
model systems.  However, as density is increased, $D_R$ becomes
increasingly underestimated by the soft-sphere relation.
Interestingly, as is shown in Fig.~\ref{fig:diff_sx_thalf},    
the quantity $DT^{-1/2}$ does approximately collapse for all densities
when plotted versus
$-\sx$ for the GC fluid.  At present, it is not known why this
alternative scaling holds for this system or whether it 
can be expected to describe the behaviors of 
other fluids with bounded potentials
(e.g., the penetrable sphere model~\cite{Lang2000Fluid-and-solid}).
We are currently carrying out a systematic investigation of the
thermodynamic and dynamic behavior of a variety of bounded potential
fluids to explore this issue, and we will 
report our findings in a future study.

In conclusion, we present molecular simulation results which
demonstrate that the GC fluid exhibits cascading regions of density, 
self-diffusivity, and
structural (two-body excess entropy) 
anomalies in the temperature-density plane.  Interestingly, an
appropriately normalized self-diffusivity of the GC fluid follows a
scaling with the two-body excess entropy that is similar to that
observed for simple atomic fluids, which do not exhibit anomalies.
The significance of this relation is that the self-diffusivity anomaly
can be viewed as simply a reflection of the structural anomalies of
the GC fluid.  An analysis of the pair correlation function shows that
the molecular origins of the structural anomalies of the GC fluid are
easy to understand and are qualitatively similar to those of 
liquid water and colloids with
short-range attractions.  Specifically, changes in thermodynamic
parameters give rise to an increase in pair correlations at short
length scales and an associated disordering at larger length scales.
For all of these fluids, the latter effect dominates, 
giving rise to the anomalous structural trends.
Finally, we show that while the Rosenfeld diffusivity
($D_R=D\rho^{1/3}T^{-1/2}$) is approximately a function of $s_2$ alone
for the GC fluid, a different combination ($DT^{-1/2}$) collapses as a
function of $\sx$.  Whether this latter observation will generally
hold for other fluids with bounded potentials is currently an open
question which deserves more careful study.

We thank J.~L. Carmer for useful discussions.   We are also grateful
to S. van Teeffelen, C. N. Likos, and H. L{\"o}wen for
bringing Refs. \onlinecite{Likos2008Cluster-forming} and
\onlinecite{Wensink2008Long-time-self-} to our attention.  Two authors
(T.M.T and J.R.E) acknowledge financial support of the National
Science Foundation (CTS-0448721 and CTS-028772, respectively). One
author T.M.T. also acknowledges support of the Welch Foundation
(F-1696) and the David and Lucile Packard Foundation.  W.P.K.
acknowledges the National Science Foundation GRF Program. J.M.
acknowledges support by the National Institute of Diabetes and
Digestive and Kidney Diseases Intramural Research Program.  The Texas
Advanced Computing Center (TACC) and University at Buffalo Center for
Computational Research provided computational resources for this
study.


\begin{thebibliography}{54}
\bibitem[{not({\natexlab{a}})}]{note_temp_dens}
\bibinfo{note}{Specifically, we study reduced temperatures $T=0.01$, $0.011$,
  $0.012$, $0.013$, $0.014$, $0.015$, $0.016$, $0.017$, $0.018$, $0.019$,
  $0.02$, $0.025$, $0.03$, $0.035$, $0.04$, $0.045$, $0.05$, $0.06$, $0.07$,
  $0.08$, $0.09$, $0.1$, $0.15$, $0.2$, $0.3$, $0.4$, $0.5$, $0.6$, $0.7$,
  $0.8$, $0.9$, $1.0$, $1.1$, $1.2$, $1.3$, $1.4$, $1.5$ and reduced densities
  $\rho=0.01$, $0.03$, $0.05$, $0.075$, $0.1$, $0.2$, $0.3$, $0.4$, $0.5$,
  $0.6$, $0.7$, $0.8$, $0.9$, $1.0$, $1.5$, $2.0$.}
\bibitem[{not({\natexlab{b}})}]{note_min_max}
\bibinfo{note}{To calculate the location of the extrema, the raw simulation
  data along a given isotherm for densities just below, at, and just above the
  simulated extrema was fit to second degree polynomials. These polynomials
  were in turn used to calculate the locations of the extrema.}
\bibitem[{not({\natexlab{c}})}]{note_tmmc}
\bibinfo{note}{These simulations require fixed values of the activity, $\xi$,
  volume, $V$, and temperature, $T$, as inputs. The activity is defined as
  $\xi=\Lambda^{-3}\exp(\mu/\kB T)$ , where $\mu$ is the chemical potential and
  $\Lambda$ is the thermal de Broglie wavelength. For the GC-TMMC simulations
  employed here, we set $V = 343$ and $\xi= 1$, and we collect information over
  a range of densities spanning from $\rho=0.0$ to $2.0$ at select values of
  the temperature. System size effects were examined by performing a subset of
  analogous calculations with $V = 646$, and were found to be negligible.
  Thermodynamic properties at values of activity other than $\xi = 1$ were
  readily obtained via the histogram reweighting technique.}
\end{thebibliography}







\end{document}